\documentclass[sigconf, dvipsnames]{acmart}

\AtBeginDocument{%
  }

\copyrightyear{2024}
\acmYear{2024}
\setcopyright{rightsretained}
\acmConference[HPDC '24]{The 33rd International Symposium on
High-Performance Parallel and Distributed Computing}{June 3--7, 2024}{Pisa,
Italy}
\acmBooktitle{The 33rd International Symposium on High-Performance Parallel
and Distributed Computing (HPDC '24), June 3--7, 2024, Pisa,
Italy}\acmDOI{10.1145/3625549.3658689}
\acmISBN{979-8-4007-0413-0/24/06}

\usepackage{bm}
\usepackage{enumitem}
\usepackage{graphicx}
\usepackage{pifont}
\usepackage{textcomp}
\usepackage{xcolor}
\usepackage{booktabs}
\usepackage{todonotes}
\usepackage{listings}
\usepackage[frozencache,cachedir=minted-cache]{minted}
\usepackage{annotate-equations}
\usepackage[most]{tcolorbox}
\usepackage{cleveref}

\definecolor{dkgreen}{RGB}{0,64,0}
\definecolor{ltgray}{RGB}{245,245,245}
\definecolor{mauve}{RGB}{139,0,139}

\definecolor{belizehole}{HTML}{2980b9}
\newtcolorbox{RQcallout}[1][]{%
  colback=black!5,
  colframe=black!5,
  notitle,
  sharp corners,
  borderline west={2pt}{0pt}{belizehole!80!black},
  enhanced,
  breakable,
}

\newcommand{\tweakedsim}{\raise.17ex\hbox{$\scriptstyle\mathtt{\sim}$}}

\newcommand{\cmark}{\ding{51}}%
\newcommand{\xmark}{\ding{55}}%

\newcommand{\generatebench}{\textsc{ParEval}}
\newcommand{\generatebenchfull}{Parallel Code Generation Evaluation}

\begin{document}

\title{Can Large Language Models Write Parallel Code?}

\author{Daniel Nichols}
\email{dnicho@umd.edu}
\orcid{0000-0002-3538-6164}
\affiliation{%
  \institution{Department of Computer Science, University of Maryland}
  \city{College Park}
  \state{Maryland}
  \country{USA}
}

\author{Joshua H. Davis}
\email{jhdavis@umd.edu}
\orcid{0000-0002-6704-0520}
\affiliation{%
  \institution{Department of Computer Science, University of Maryland}
  \city{College Park}
  \state{Maryland}
  \country{USA}
}

\author{Zhaojun Xie}
\email{zxie12@umd.edu}
\orcid{0009-0004-2952-4198}
\affiliation{%
  \institution{Department of Computer Science, University of Maryland}
  \city{College Park}
  \state{Maryland}
  \country{USA}
}

\author{Arjun Rajaram}
\email{arajara1@umd.edu}
\orcid{0000-0001-7016-1617}
\affiliation{%
  \institution{Department of Computer Science, University of Maryland}
  \city{College Park}
  \state{Maryland}
  \country{USA}
}

\author{Abhinav Bhatele}
\email{bhatele@cs.umd.edu}
\orcid{0000-0003-3069-3701}
\affiliation{%
  \institution{Department of Computer Science, University of Maryland}
  \city{College Park}
  \state{Maryland}
  \country{USA}
}

\renewcommand{\shortauthors}{Nichols et al.}

\newcommand{\RQone}{How well do state-of-the-art LLMs generate parallel
  code, and which models are the best?}
\newcommand{\RQoneAnswer}{We show that all tested LLMs, both open- and 
  closed-source, struggle to generate parallel code. Of the models tested,
  GPT-3.5 performs the best with a pass@1 of 76.0 for serial code generation
  and a pass@1 of 39.6 for parallel code generation.}
\newcommand{\RQtwo}{Which parallel execution models and problem types are most
  challenging for LLMs?}
\newcommand{\RQtwoAnswer}{We observe that LLMs struggle most with MPI 
  code generation, and perform best for OpenMP and Kokkos code generation.
  Additionally, we show that LLMs find it challenging to generate parallel
  code for sparse, unstructured problems.}
\newcommand{\RQthree}{How performant and scalable is the parallel code
  generated by LLMs?}
\newcommand{\RQthreeAnswer}{We observe that the parallel code generated by LLMs
  can have poor parallel speedup and efficiency. Additionally, we show that the
  LLMs that most often generate correct parallel code do not necessarily
  generate the most performant parallel code.}
\newcommand{\RQfour}{How well can LLMs translate between execution
  models? How performant and scalable is the translated code?}
\newcommand{\RQfourAnswer}{We show that providing LLMs with correct 
  implementations in one execution model can improve their ability to
  generate correct code in another execution model. This is particularly true
  for smaller open-source models.}

\begin{abstract}
Large language models are increasingly becoming a popular tool for software
development. Their ability to model and generate source code has been
demonstrated in a variety of contexts, including code completion,
summarization, translation, and lookup. However, they often struggle to
generate code for complex programs. In this paper, we study the capabilities of
state-of-the-art language models to generate parallel code. In order to
evaluate language models, we create a benchmark, \generatebench{}, consisting
of prompts that represent 420 different coding tasks related to scientific and
parallel computing. We use \generatebench{} to evaluate the effectiveness of
several state-of-the-art open- and closed-source language models on these
tasks. We introduce novel metrics for evaluating the performance of generated
code, and use them to explore how well each large language model performs for
12 different computational problem types and six different parallel programming
models. 

\end{abstract}



\maketitle

\section{Introduction}
\label{sec:introduction}
Large language model (LLM) based coding tools are becoming popular in software
development workflows. Prior work has demonstrated their effectiveness at
performing a variety of tasks, including code completion, summarization,
translation, and lookup~\cite{Gu2022AssembleFM, Ahmed2022LearningCS,
Haque2022SemanticSM, Ahmad2020ATA, Richter2022CanWL, Kharkar2022LearningTR,
Garg2022DeepDevPERFAD}.  Popular models such as
StarCoder~\cite{li2023starcoder}, span a wide range of programming languages
and domains, and can be used to complete or generate code during the
development process. This makes them a promising tool for improving developer
productivity and the overall quality of software. However, despite the rapid
advancement and scaling of LLMs in recent years, they still struggle with more
complicated tasks such as reasoning and planning. One particularly complex task
that LLMs struggle with is generating {\em parallel} code. This task involves
reasoning about data distributions, parallel algorithms, and parallel
programming models.

Parallel code is essential to modern software development due to the ubiquity of
multi-core processors, GPGPUs, and distributed systems. However, writing
parallel code is difficult and error-prone. Parallel algorithms are generally
more complicated than their sequential counterparts, and parallel bugs such as
race conditions and deadlocks are notoriously non-trivial to debug.  Further, it
can be challenging to reason about the performance of parallel code and identify
``performance bugs''~\cite{azad:msr2023}. LLMs can potentially help developers
overcome these challenges but, this requires an understanding of the current
capabilities of LLMs, and in turn, a well-designed and reproducible methodology
to assess these capabilities.

There are several existing benchmarks for evaluating the capabilities of LLMs to
generate correct code. However, none of them test generation of {\it parallel}
code. Most existing benchmarks focus on short, array or string manipulation
tasks, and are predominantly in Python (or translated to other languages from
Python~\cite{multiple}). Only more recent benchmarks such as
DS-1000~\cite{lai2022ds1000}, test the usage of APIs, which are critical to
using parallel programming models. Further, these benchmarks do not evaluate the
performance of the generated code, instead testing only correctness. While
correctness is a crucial metric, performance is also vital for developers
writing parallel code. Thus, it is imperative to design new benchmarks and
metrics to evaluate the usefulness of LLMs for parallel code generation tasks.

Developing a set of benchmarks that fully covers the space of desired
capabilities is non-trivial. Identifying the best LLM for parallel code
generation requires testing on problems that cover shared- and
distributed-memory programming models, different computational problem types,
and different parallel algorithms. This can become a large quantity of
benchmarks that must be manually designed. Further, these benchmarks are
challenging to test. Traditional Python code generation benchmarks are tested
by running {\it eval} on the generated code for a small number of small unit
tests.  On the other hand, in the case of parallel code --- we must compile
C/C++ code, link against one or more parallel libraries, and run the code in
the proper parallel environment.  Additionally, if we want to test the
performance of the generated code, then we must choose reasonable input sizes
for each benchmark.

In order to evaluate the current capabilities and limitations of LLMs in
generating parallel code, we propose the \generatebenchfull{}
(\generatebench{}) benchmark: a set of benchmarks (prompts) for evaluating how
well LLMs generate parallel code. These benchmarks cover twelve different
computational problem types, and seven different execution models: serial,
OpenMP, Kokkos, MPI, MPI+OpenMP, CUDA, and HIP. We evaluate several
state-of-the-art open- and closed-source LLMs using these benchmarks, and
report metrics that represent the {\it correctness} and {\it performance} of
the generated code. We introduce novel code generation evaluation metrics that
assess performance and parallel scaling. We further analyze how each model
performs with respect to the various programming models and computational
problem types. We discuss the areas where current state-of-the-art LLMs are
already performing well and the areas where they can be improved.

In this paper, we make the following important contributions:
\begin{itemize}
    \item We design the \generatebench{} benchmark for evaluating the ability
of LLMs to generate and translate parallel code. \generatebench{} is available online at:
\href{https://github.com/parallelcodefoundry/ParEval}{github.com/parallelcodefoundry/ParEval}.
    \item We introduce two novel metrics, speedup$_n$@k and \\ efficiency$_n$@k, 
        for evaluating the performance and scaling of LLM generated code.
    \item We evaluate the effectiveness of several state-of-the-art open- and
        closed-source LLMs using the \generatebench{} benchmark.
    \item We identify several areas where current state-of-the-art LLMs can
        improve their capabilities on parallel code generation.
\end{itemize}

In addition to these contributions, we explore the following research
questions (answers based on our observations):
\begin{itemize}
    \item[{\bf RQ1}] {\it \RQone{}} \RQoneAnswer{}
    \item[{\bf RQ2}] {\it \RQtwo{}} \RQtwoAnswer{}
    \item[{\bf RQ3}] {\it \RQthree{}} \RQthreeAnswer{}
    \item[{\bf RQ4}] {\it \RQfour{}} \RQfourAnswer{}
\end{itemize}

\section{Background}
\label{sec:background}
In this section, we provide background information on large language models and
how they are used for text generation. We further discuss how large language
models can be used for code generation.

\subsection{Large Language Models}\label{sec:bg-llms}

Natural Language Processing (NLP) has largely been dominated by
transformer-based models since their introduction in 2017 by Vaswani et
al.~\cite{transformer}. Transformer networks are designed to model sequential
data, such as text, relying on {\it self-attention} mechanisms to model the
relationships between values in a sequence. Self-attention enables modeling of
long-range dependencies in the data without vanishing gradient and scaling
issues and allows for sequence elements to processed in parallel. Transformers
learn {\it attention scores}, which are computed between pairs of tokens in the
input. {\it Multi-head attention} allows for learning multiple attention
representations. These large transformer models are generally trained to model
the distribution of a text corpus such as the English language by predicting
the next token in a sequence given previous tokens. Transformer-based models
have emerged as the most effective means of modeling text data, and have been
shown to be effective at a wide range of NLP tasks.

\subsection{Large Language Models for Code}\label{sec:bg-llms-code}

An LLM trained on a large corpus of code can be used to generate code by giving
it a code input prompt and asking it to predict the next token. Generally, code
LLMs are trained on a large corpus of code, such as The
Stack~\cite{Kocetkov2022TheStack}, that covers a wide range of programming
languages and application types. Sometimes the pre-training corpus includes
natural language as well, such as The Pile~\cite{the_pile,
xu_2022_code-llms-survey-dataset}. In some instances, such as
CodeLlama~\cite{roziere2023code}, the code LLM is a natural language model that
has been further fine-tuned on a corpus of code. When generating code with one
of these models it is often not enough to simply select the most probable next
token to construct a sequence. This often leads to repetitive, low-quality
outputs~\cite{holtzman:iclr2020}, so we also need a strategy for token
selection. We utilize {\it nucleus sampling} and {\it model temperature} in
this study.

\subsubsection*{Nucleus Sampling}\label{sec:bg-nucleus}

Nucleus sampling~\cite{holtzman:iclr2020}, also called {\it top-p sampling},
samples the next token from the token probability distribution up to some
cut-off $p$ in the cumulative distribution function. Compared to sampling from
a fixed number of top tokens in the distribution (called {\it top-k sampling}),
this ensures the selection of a more representative sample of tokens from the
distribution. Nucleus sampling is often used in code generation tasks with a
value of $p=0.95$ and is sometimes combined with top-$k$ sampling.

\subsubsection*{Model Temperature}\label{sec:bg-temperature}

Generation {\it temperature} is a scaling value applied to the raw model
outputs, or {\it logits}, before they are converted to a probability
distribution.  The value is applied by first dividing the logits vector by the
scalar temperature before computing the softmax of the logits. Higher
temperatures make the probability distribution more peaked, upweighting the
most probably tokens, while lower temperatures make the distribution more
uniform.  Intuitively, lower temperatures yield more conservative generations
that the model is more {\it confident} in. Conversely, higher temperatures will
lead to more varied and {\it creative} generations. For code generation tasks,
a low temperature value of $0.2$ is often used.

\section{Related Work}
\label{sec:related-work}
Below, we describe related work in benchmarking LLMs for code-related tasks
and applying LLMs to parallel and HPC code.

\subsection{Benchmarking LLMs for Code-related Tasks}

Since the introduction of the Codex model and HumanEval
benchmark~\cite{codex-copilot-all-author}, many works have proposed new LLMs
for code and evaluated them on a variety of tasks. The number of code-specific
models has grown rapidly as open-source models and data sets become more
available and low-rank training techniques, such as LoRA~\cite{lora}, make
training large models more feasible. These models are usually evaluated on code
generation tasks such as HumanEval~\cite{codex-copilot-all-author},
MBPP~\cite{mbpp}, and DS-1000~\cite{lai2022ds1000}.

The first of these, HumanEval~\cite{codex-copilot-all-author}, is a set of 164
code generation tasks that are designed to evaluate the ability of LLMs to
write short Python functions that solve a variety of problems, given a
docstring and function signature.  Similar to HumanEval is the Mostly Basic
Python Problems (MBPP)~\cite{mbpp} benchmark which is a set of 1000 simple
Python problems.  MBPP is often evaluated with few-shot prompts, where example
correct solutions to other problems are included in the prompts. A common
extension of these benchmarks is MultiPL-E~\cite{multiple} which extends the
set of HumanEval and MBPP tests to 18 programming languages.

The DS-1000 benchmark~\cite{lai2022ds1000} tests the ability of the model to
generate more complex, data science-related code, for 1000 tasks making use of
common data science libraries. Other similar benchmarks that evaluate coding
LLMs on more complex tasks are GSM8K~\cite{cobbe2021gsm8k} and
GSM-HARD~\cite{gao2022pal}, which use PAL~\cite{gao2022pal} to evaluate the
ability of LLMs to generate Python code snippets to assist in chains of
reasoning.  The CoderEval benchmarks~\cite{yu2023codereval} are a set of 230
Java and 230 Python code generation tasks that require the model to write
context-dependent functions, rather than standalone functions as in HumanEval
and MBPP.

Additionally, there have been several domain specific benchmarks that evaluate
more narrow uses of LLM code
generation~\cite{liu2023verilogeval,du2023classeval,tang2023biocoder}. All of
these benchmarks make use of tasks manually created by experts to test more
specific use cases of LLMs.

\subsection{Applying LLMs to Parallel and HPC Code}

Recently there has been a growing interest in applying LLMs to parallel and
High Performance Computing (HPC) code. Several works have looked at creating
smaller specialized HPC models~\cite{nichols:arxiv2023, kadosh2023scope} or
applying existing LLMs to HPC tasks~\cite{munley2023llm4vv, chen2023data,
chen2023lm4hpc}.  Nichols et al.~\cite{nichols:arxiv2023} introduce HPCCoder, a
model fine-tuned on HPC code, and evaluate its ability to generate HPC code,
label OpenMP pragmas, and predict performance. Kadosh et
al.~\cite{kadosh2023scope} introduce TOKOMPILER, an HPC specific tokenizer for
LLMs, and use it to train COMPCODER, a model trained on C, C++, and Fortran
code.

Other works have looked at applying existing LLMs to HPC tasks. Munley et
al.~\cite{munley2023llm4vv} evaluate the ability of LLMs to generate compiler
verification tests for parallel OpenACC code. Chen et al.~\cite{chen2023data}
use LLMs to identify data races in parallel code and propose the DRB-ML data
set, which is integrated into the LM4HPC framework~\cite{chen2023lm4hpc}. Godoy
et al.~\cite{Godoy_2023} and Valero-Lara et al.~\cite{valerolara2023comparing}
both evaluate the capabilities of LLMs on generating HPC kernels, but use a
limited set of problems and LLMs and do not prompt or evaluate the LLMs using
standard practices. None of these works comprehensively evaluate and compare the
ability of LLMs to generate parallel code across a large number of problems,
execution models, and LLMs using state-of-the-art evaluation techniques, which
is the focus of this work.

\section{\generatebench{}: Prompts for Parallel Code Generation}
\label{sec:benchmarks}

In order to evaluate the ability of LLMs to generate parallel code, we propose
the \generatebenchfull{} (\generatebench{}) benchmark. Below, we discuss the
design of \generatebench{}, and its various components that lead to the creation
of concrete prompts for LLMs.

To disambiguate the use of the terms {\it prompt}, {\it task}, {\it problem},
{\it problem type}, and {\it benchmark} we define them as follows.
\begin{itemize}[leftmargin=*]
    \item {\it Task/Prompt}: An individual text prompt that is given to the
        LLM to generate code. The output can be compiled, executed, and scored
        as either correct or incorrect code.
    \item {\it Problem}: A set of tasks or prompts that test the ability of the LLM
        to generate code for the same computational work, but each task
        or prompt may use a different execution model.
    \item {\it Problem Type}: A set of problems that test computational 
        problems with similar work or from similar domains (for example, {\it sorting} 
        problems).
    \item {\it Benchmark}: A set of prompts that are all tested together to
        evaluate the performance of the LLM. We name the collection of all the prompts we have designed as the \generatebench{}
        benchmark.
\end{itemize}

\subsubsection*{\bf Benchmark Requirements}\label{sec:benchmark-requirements}

The goal of \generatebench{} is to evaluate the ability of LLMs to generate
parallel code. To do this, the prompts should be such that:
\begin{enumerate}[leftmargin=*]
\item The prompts should cover a wide variety of computational
problem types, and parallel programming models.
\item The prompts should be simple enough that
they can be generated as a standalone function, but complex enough that they
are not too trivial to solve. 
\item The prompts should not exist
within any of the LLMs' training datasets, to prevent the LLMs from simply copying
solutions from their training data.
\item The prompts and corresponding outputs should be able to be evaluated
automatically, since there will be many different tasks and LLM outputs.
\end{enumerate}

\vspace{0.1in}
In order to fulfill the requirements above, we propose \generatebench{}, a set of 420 prompts that cover
twelve different computational problem types and seven different execution
models.  Each problem type has five different problems, and each problem has a
prompt for each of the seven execution models, resulting in 420 total prompts.
Each prompt in \generatebench{} is a standalone function that requires the LLM
to generate code that solves the problem either sequentially or in parallel.

\subsubsection*{\bf Problem Types}\label{sec:problem-types}

The problem types are listed and described in ~\Cref{tab:benchmark-overview}.
These were hand-selected by us, and represent a wide variety of common
computational problems that are often parallelized. Each requires different
strategies or APIs to solve in parallel. For instance, the problems in the {\it
Sort} problem type require the LLM to generate code that sorts an array of
values.

\begin{table}[h]
    \centering
    \caption{Descriptions of the twelve problem types in \generatebench{}. Each
problem type has five concrete problems, and each problem has a prompt for all
seven execution models.}
    \label{tab:benchmark-overview}
    \begin{tabular}{@{}ll@{}}
    \toprule
    \textbf{Problem Type} & \textbf{Description}                                                                                              \\ \midrule
    Sort                  & \begin{tabular}[c]{@{}l@{}}Sort an array or sub-array of values;\\ in-place and out-of-place.\end{tabular}        \\
    Scan                  & \begin{tabular}[c]{@{}l@{}}Scan operations, such as prefix sum, \\ over an array of values.\end{tabular}          \\
    Dense Linear Algebra  & \begin{tabular}[c]{@{}l@{}}Dense linear algebra functions from\\ all three levels of BLAS.\end{tabular}               \\
    Sparse Linear Algebra & \begin{tabular}[c]{@{}l@{}}Sparse linear algebra functions from\\ all three levels of BLAS.\end{tabular}              \\
    Search                & \begin{tabular}[c]{@{}l@{}}Search for an element or property in\\ an array of values.\end{tabular}                \\
    Reduce                & \begin{tabular}[c]{@{}l@{}}Reduction operation over an array \\ dimension, such as computing a sum.\end{tabular}  \\
    Histogram             & \begin{tabular}[c]{@{}l@{}}Binning values based on a\\ property of the data.\end{tabular}                         \\
    Stencil               & \begin{tabular}[c]{@{}l@{}}One iteration of 1D and 2D stencil \\ problems, such as Jacobi relaxation.\end{tabular}     \\
    Graph                 & \begin{tabular}[c]{@{}l@{}}Graph algorithms, such as \\ component counting.\end{tabular}                          \\
    Geometry              & \begin{tabular}[c]{@{}l@{}}Compute geometric properties,\\ such as convex hull.\end{tabular}                      \\
    Fourier Transform     & \begin{tabular}[c]{@{}l@{}}Compute standard and inverse\\ Fourier transforms.\end{tabular}                        \\
    Transform             & \begin{tabular}[c]{@{}l@{}}Map a constant function to each\\ element of an array.\end{tabular}                    \\ \bottomrule
    \end{tabular}
\end{table}

\subsubsection*{\bf Problems}\label{sec:problems}

The five problems within each problem type are designed to test the core
functionality of the problem type.  To prevent prompting the model for a
solution that is already in its training dataset, the five problems are small
variations of the usual problem type.  For example, one of the scan problems is
to compute the {\it reverse} prefix sum of an array, rather than directly
computing the prefix sum. These variations still test the model's understanding
of the core computational problem, but mitigate the likelihood of it simply
copying code from its training dataset. \Cref{lst:example-kokkos-prompt}
shows another example of these problem variations. Another benefit of having
five problems per problem type is that it provides more data points for
evaluating the LLM's performance on that problem type, but not so many that it
becomes infeasible to implement and maintain.

\subsubsection*{\bf Prompts}\label{sec:prompts}

Each problem has a prompt for each of the seven execution models that the LLM
is required to generate code for.  The seven execution models we test are:
serial, OpenMP~\cite{OpenMP4}, MPI~\cite{snir1998mpi}, MPI+OpenMP,
Kokkos~\cite{kokkos:tpds2022}, CUDA~\cite{cuda}, and HIP~\cite{hip}. All the
prompts are in C++, CUDA, or HIP.  These represent both shared and distributed
memory programming models, as well as GPU programming models. The prompts for
each execution model are designed to be as similar to the other prompts for
that problem as possible, while still being idiomatic for the programming
model. For serial, OpenMP, MPI, and MPI+OpenMP prompts, we use STL data
structures such as \verb|std::vector| and \verb|std::array|. For Kokkos, we
utilize the \verb|Kokkos::View| data structure (as shown in
\Cref{lst:example-kokkos-prompt}). The CUDA and HIP prompts use raw pointers to
represent array structures.

\begin{listing}[ht]
\begin{minted}[fontsize=\small]{cpp}
#include <Kokkos_Core.hpp>

/* Replace the i-th element of the array x with the minimum
   value from indices 0 through i.
   Use Kokkos to compute in parallel. Assume Kokkos has
   already been initialized.
   Examples:

   input: [8, 6, -1, 7, 3, 4, 4]
   output: [8, 6, -1, -1, -1, -1, -1]

   input: [5, 4, 6, 4, 3, 6, 1, 1]
   output: [5, 4, 4, 4, 3, 3, 1, 1]
*/
void partialMinimums(Kokkos::View<float*> &x) {
\end{minted}
\caption{An example {\it Scan} prompt for Kokkos. The LLM will be tasked
    with completing the function body.}
\label{lst:example-kokkos-prompt}
\end{listing}

We list an example prompt in \Cref{lst:example-kokkos-prompt} for a
variant of a scan problem to generate Kokkos code. The goal of this problem is to compute the minimum
value of the array up to each index. We include example inputs and outputs in
the prompt as this can significantly improve the quality of the generated
code~\cite{mbpp}. The necessary \verb|#include| statements are also prepended
to the prompt as we found that this improves the likelihood of the LLM
correctly using the required programming model.

\begin{table*}[t]
    \caption{The models compared in our evaluation. CodeLlama and its variants
currently represent state-of-the-art open-source LLMs and GPT represents
closed-source LLMs. OpenAI does not publish the numbers of parameters in their
models.}
    \label{tab:models}
    \centering
    \begin{tabular}{@{}lrcccc@{}}
    \toprule
    \textbf{Model Name}                                    & \begin{tabular}[c]{@{}c@{}}{\bf No.~of}\\{\bf Parameters}\end{tabular} & \begin{tabular}[c]{@{}c@{}}{\bf Open-source}\\{\bf Weights}\end{tabular} & \textbf{License}   & \begin{tabular}[c]{@{}c@{}}{\bf HumanEval}\textsuperscript{\dag} \\ (pass@1)\end{tabular}    & \begin{tabular}[c]{@{}c@{}}{\bf MBPP}\textsuperscript{\ddag}\\ (pass@1)\end{tabular}    \\ \midrule
    CodeLlama-7B~\cite{roziere2023code}              & 6.7B                      & \cmark                     & llama2             & 29.98                                                                                        & 41.4                                                                                    \\
    CodeLlama-13B~\cite{roziere2023code}             & 13.0B                     & \cmark                     & llama2             & 35.07                                                                                        & 47.0                                                                                    \\
    StarCoderBase~\cite{li2023starcoder}             & 15.5B                   & \cmark                     & BigCode OpenRAIL-M & 30.35                                                                                        & 49.0                                                                                    \\
    CodeLlama-34B~\cite{roziere2023code}             & 32.5B                     & \cmark                     & llama2             & 45.11                                                                                        & 55.0                                                                                    \\
    Phind-CodeLlama-V2~\cite{phind-codellama-34b-v2} & 32.5B                     & \cmark                     & llama2             & 71.95                                                                                        & ---                                                                                     \\
    GPT-3.5~\cite{gpt-3}                             & ---                     & \xmark                     & ---                & 61.50                                                                                        & 52.2                                                                                    \\
    GPT-4~\cite{openai2023gpt4}                      & ---                     & \xmark                     & ---                & 84.10                                                                                        & ---                                                                                     \\ \bottomrule
    \multicolumn{6}{@{}l@{}}{\footnotesize \dag HumanEval results are from the BigCode Models Leaderboard~\cite{bigcode_leaderboard}, except for GPT-3.5 and GPT-4 which are from~\cite{zeroshotreplication}.} \\
    \multicolumn{6}{@{}l@{}}{\footnotesize \ddag MBPP results are from~\cite{roziere2023code}.} \\
    \end{tabular}
\end{table*}

\section{Description of Evaluation Experiments}
\label{sec:tasks}
Now that we have described the prompts in the previous section, we describe how
we can use them to evaluate the performance of LLMs on two different tasks --
code generation and translation.

\subsection{Experiment 1: Parallel Code Generation}\label{sec:setup-code-generation}

The first experiment studies the ability of LLMs to {\em generate} code, either
sequential or in a specific parallel programming model, given a simple
description in a prompt (see~\Cref{lst:example-kokkos-prompt}). We evaluate LLMs
on how well they can generate code for all the prompts in \generatebench{}. We
do so by asking the model to complete the function started in the prompt, and
then evaluating the generated code. By compiling and executing the generated
code, we report different metrics that will be described in
Section~\ref{sec:metrics}. The metrics are computed over the combined results
from the OpenMP, MPI, MPI+OpenMP, Kokkos, CUDA, and HIP execution models, and
compared with the same metrics computed over the serial results. These results
will provide insight into how well the model can write parallel code based on
natural language descriptions. The results can also be compared along the axes
of execution model and problem type.

\subsection{Experiment 2: Parallel Code Translation}\label{sec:setup-code-translation}

The second experiment studies the ability of LLMs to effectively {\em translate}
code provided in one execution model to another execution model. To accomplish
this, we prompt the LLM with a correct version of the code in one execution
model and ask it to translate it to another execution model. An example of this
prompt format is shown in \Cref{lst:translation-prompt}. We evaluated several
prompting formats for translation, such as giving examples of other successful
translations, but found the format in \Cref{lst:translation-prompt} to be the
most effective.

\begin{listing}[ht]
\begin{minted}[fontsize=\footnotesize]{cpp}
// A serial implementation of sumOfMinimumElements
/* Return the sum of the minimum value at each index of vectors 
   x and y for all indices.
   i.e. sum = min(x_0, y_0) + min(x_1, y_1) + min(x_2, y_2) + ...
   Example:

   input: x=[3, 4, 0, 2, 3], y=[2, 5, 3, 1, 7]
   output: 10
*/
double sumOfMinimumElements(std::vector<double> const& x, 
    std::vector<double> const& y) {
    double sum = 0.0;
    for (size_t i = 0; i < x.size(); ++i) {
        sum += std::min(x[i], y[i]);
    }
    return sum;
}

// An OpenMP implementation of sumOfMinimumElements
/* Return the sum of the minimum value at each index of vectors 
   x and y for all indices.
   i.e. sum = min(x_0, y_0) + min(x_1, y_1) + min(x_2, y_2) + ...
   Use OpenMP to sum in parallel.
   Example:

   input: x=[3, 4, 0, 2, 3], y=[2, 5, 3, 1, 7]
   output: 10
*/
double sumOfMinimumElements(std::vector<double> const& x, 
   std::vector<double> const& y) {
\end{minted}
\caption{An example prompt given to the model for code translation. The model
   is given a sequential implementation of sumOfMinimumElements and tasked with
   translating it to OpenMP.}
\label{lst:translation-prompt}
\end{listing}

In theory, we could have evaluated translation capabilities between each pair of
execution models for each problem. However, to limit the quadratic increase in
the number of prompts, we only evaluate translations for these pairs:
$\mathrm{serial}\rightarrow\mathrm{OpenMP}$,
$\mathrm{serial}\rightarrow\mathrm{MPI}$, and
$\mathrm{CUDA}\rightarrow\mathrm{Kokkos}$. We identify these as some of the most
relevant translation tasks for HPC developers. We compute the same metrics as
for Experiment 1. These results will provide insight into how well the model can
translate between different execution models. The results can also be compared
along the axes of source and target execution model and problem type.

\section{Models used for Comparison}
\label{sec:models}
We choose to compare several state-of-the-art open-source and closed-source
LLMs, as well as smaller LLMs that are more practical for use in production.
We provide brief descriptions of the LLMs used in our evaluation, and their
properties below. \Cref{tab:models} provides a summary and some salient
properties of the models used.

\subsubsection*{\bf CodeLlama (CL-7B, CL-13B, and CL-34B)}\label{sec:models-codellama}

Rozière et al.~originally introduced CodeLlama models in~\cite{roziere2023code}
as variants of the Llama 2 model~\cite{touvron2023llama}, fine-tuned for code.
All three models started with Llama 2 weights and were then fine-tuned on 500
billion tokens from a dataset of predominantly code. The Llama 2 models were
also extended to support longer context lengths of 16k and infilling to
generate code in the middle of sequences. We select these models as they are
amongst the top performing open-source LLMs. Additionally, the CodeLlama models
are very accessible as there are small model sizes available and there exists a
thriving software ecosystem surrounding Llama 2 based models.

\subsubsection*{\bf StarCoderBase}\label{sec:models-starcoder}

The StarCoderBase model~\cite{li2023starcoder} is a 15.5B parameter model
trained on 1 trillion tokens from The Stack~\cite{Kocetkov2022TheStack}. In
addition to code from 80+ programming languages, its data set includes natural
language in git commits and Jupyter notebooks. StarCoderBase supports infilling as
well as a multitude of custom tokens specific to code text data. The model
architecture is based on the SantaCoder model~\cite{allal2023santacoder}, and it
supports a context length of 8K tokens. We select StarCoderBase as it is one of the
best performing open-source models around its size, and is frequently used for
comparisons in related literature.

\subsubsection*{\bf Phind-CodeLlama-V2}\label{sec:models-phind}

The Phind-CodeLlama-V2 model~\cite{phind-codellama-34b-v2} is a CodeLlama-34B
model fine-tuned on over 1.5 billion tokens of code data. At the time we were
selecting models for comparison it topped the BigCode Models
Leaderboard~\cite{bigcode_leaderboard} among open-access models on HumanEval
with a pass@1 score of 71.95. However, the fine-tuning dataset for this model is
not publicly available, so it is not possible to ensure that the BigCode
benchmarks themselves are not included in Phind's fine-tuning dataset.

\subsubsection*{\bf GPT-3.5 and GPT-4}\label{sec:models-gpt}

GPT-3.5 and GPT-4 are closed-source LLMs from
OpenAI~\cite{gpt-3,openai2023gpt4}. Most information about these models is not
publicly available, however, they can be used for inference via a paid API. We
use the most up-to-date versions of these models available at the time of
writing, the {\it gpt-3.5-turbo-1106} and {\it gpt-4-1106-preview} models.
Unlike the other models tested, these are instruction-tuned and aligned to
human preferences. Rather than using them for direct code generation, we have to 
interact with them via a chat interface. As with the Phind-CodeLlama-V2 model, the
data used to train these models is not publicly available, so it is difficult
to fairly compare them with the other models as they might have seen some
prompts during training.

\section{Evaluation Metrics}
\label{sec:metrics}
It is important to be able to meaningfully compare the performance of the
selected LLMs at generating correct and efficient code for the prompts in
\generatebench{}. This section details how we accomplish this by adopting a
popular correctness metric for code LLMs, and defining two new
performance-related metrics.

\subsection{Metric for Correctness}\label{sec:metrics-correctness}

We adopt the pass@$k$ metric from~\cite{codex-copilot-all-author} to quantify
correctness of the generated code. For a given prompt, pass@$k$ estimates the
probability that the model will generate a correct solution given $k$ attempts.
Often the average pass@$k$ over all prompts in a benchmark is reported. To
estimate the pass@$k$ over a set of prompts, we first generate $N$ samples for
each prompt using the model, where $N>k$. These samples are then evaluated for
correctness. The number of correct samples can be used to estimate the pass@$k$
value as shown in \Cref{eq:pass-k}.

\vspace{1em}
\begin{equation}\label{eq:pass-k}
    \text{pass@}k =
    \frac{1}{\lvert \eqnmarkbox[MidnightBlue]{P1}{P}\rvert}
    \sum_{p\in \eqnmarkbox[MidnightBlue]{P2}{P}}
    \left[
        1 -
        \binom{
            \eqnmarkbox[WildStrawberry]{N1}{N} -
            \eqnmarkbox[OliveGreen]{cp}{c_p}
        }{
            k
        }
        /
        \binom{
            \eqnmarkbox[WildStrawberry]{N2}{N}
        }{
            k
        }
    \right]
\end{equation}
\annotatetwo[yshift=1em]{above}{N1}{N2}{Number of samples generated per prompt}
\annotatetwo[yshift=-1em]{below}{P1}{P2}{Set of prompts}
\annotate[yshift=-2.3em]{below,right}{cp}{Number of correct\\samples for prompt $p$}
\vspace{1.5em}

This metric provides insight into how often do models generate correct
code. The probability that the model will generate a correct solution in one
attempt, pass@$1$, is the most useful metric for end-users as it aligns with
how LLMs are used in practice. In this paper, we report
$100 \times \mathrm{pass}@k$ as is common in related literature and online
leaderboards~\cite{bigcode_leaderboard,codex-copilot-short-author}.
Additionally, as models have become more capable, studies have shifted toward
only reporting pass@$1$ values. However, pass@$k$ values for $k>1$ are still
useful for understanding how models perform on more difficult prompts. Commonly
reported values of $k$ are 1, 5, 10, 20, and 100. It is also common to
report pass@$1$ values using a generation temperature of 0.2 and pass@$k$ for
higher values of $k$ using a generation temperature of 0.8. This higher
temperature allows the model to more extensively explore the solution space
when generating a larger number of attempts.

\subsection{Performance Metrics}\label{sec:metrics-performance}

For parallel and HPC code, it is important to consider both the correctness and
performance of the generated code. To analyze and compare the runtime performance of LLM generated code, we introduce two new metrics: $\mathrm{speedup}_n@k$ and
$\mathrm{efficiency}_n@k$.

\subsubsection*{$\mathbf{speedup_n@k}$}

The first metric, $\mathrm{speedup}_n@k$, measures the expected best performance speedup
of the generated code relative to the performance of a sequential baseline (see \Cref{sec:setup-evaluation})
if the model is given $k$ attempts to generate the code. The relative speedup is computed based on the execution time obtained using $n$ processes or threads. For a given prompt
$p$, the expected best speedup relative to a sequential baseline, $T^*_p$,
is given by \Cref{eq:expected-perf-improvement}.

\vspace{1.5em}
\begin{equation}\label{eq:expected-perf-improvement}
    \mathbb{E}
    \left[
        \max
        \left\{
        \frac{T^*_p}{T_{p,s_1,n}}, \ldots, \frac{T^*_p}{T_{p,s_k,n}}
        \right\}
    \right]
    =
    \sum_{j=1}^N
    \frac{
        \binom{
            j-1
        }{
            k-1
        }
    }{
        \binom{
            N
        }{
            k
        }
    }
    \frac{
        \eqnmarkbox[SeaGreen]{Tsp}{T^*_p}
    }{
        \eqnmarkbox[Melon]{Tpj}{T_{p,j,n}}
    }
\end{equation}
\annotate[yshift=-0.5em]{below,left}{Tpj}{runtime of sample $j$ of prompt $p$
    on $n$ resources}
\annotate[yshift=1em]{above,left}{Tsp}{runtime of sequential baseline for
    prompt $p$}
\vspace{1.5em}

To demonstrate that \Cref{eq:expected-perf-improvement} represents the desired
quantity, consider the set of $N$ generated samples is in order from slowest to
fastest. This is without loss of generality as we assume the $k$ samples are
selected uniformly and, thus, all size $k$ permutations are equally likely. The
probability that the max is the $j$th sample is given by
$\binom{j-1}{k-1}/\binom{N}{k}$, as there must be $j-1$ elements before $j$
and, thus, $\binom{j-1}{k-1}$ ways to select the remaining elements. The
sum of these probabilities, each weighted by their respective speedups,
gives the expected max speedup over $k$ samples.  Taking the average of
\Cref{eq:expected-perf-improvement} over all prompts we can define the
$\mathrm{speedup}_n@k$ metric as shown in \Cref{eq:speedup-k}.
\begin{equation}\label{eq:speedup-k}
    \textrm{speedup}_n@k =
    \frac{1}{\lvert P\rvert}
    \sum_{p\in P}
    \sum_{j=1}^N
    \frac{
        \binom{
            j-1
        }{
            k-1
        }
    }{
        \binom{
            N
        }{
            k
        }
    }
    \frac{
        T^*_p
    }{
        T_{p,j,n}
    }
\end{equation}

For a single LLM, the $\mathrm{speedup}_n@k$ metric can be used to understand
how well its generated code performs compared to sequential baselines. A value
greater than 1 indicates that the generated code is faster than the baseline on
average, while a value less than 1 indicates that the generated code is
generally slower than the baseline. When comparing multiple LLMs, a higher value
of $\mathrm{speedup}_n@k$ signifies more performant code. It is
important to note that this metric is hardware dependent and, thus, to compare
models fairly all the run times need to be collected on the same hardware.

The $\mathrm{speedup}_n@k$ metric also gives insight into how well the
generated code makes use of parallelism in its computation. It is fixed to a
given number of resources, $n$, which can either be threads or
processes, depending on the model of parallelism being used. It also adds
another axis to vary when comparing models. When studying a single model, the
$\mathrm{speedup}_n@k$ metric can be compared at different values of $n$ to
understand the complete scaling behavior of that model. When comparing multiple
models, it is typically most useful to fix $n$ to a single value. One could
also average over many values of $n$, but this risks hiding too much
information to be useful.

\subsubsection*{$\mathbf{speedup_{\mathbf{max}}@k}$}

We also define a variant of the $\mathrm{speedup}_n@k$ metric,
$\mathrm{speedup}_{\mathrm{max}}@k$, as shown in~\Cref{eq:speedup-max-k}, which
estimates the maximum speedup over all $n$ and not a fixed resource count.
\begin{equation}\label{eq:speedup-max-k}
    \textrm{speedup}_{\mathrm{max}}@k =
    \frac{1}{\lvert P\rvert}
    \sum_{p\in P}
    \sum_{\substack{j=1\\n\in \mathrm{procs}}}^{N \cdot \lvert \mathrm{procs} \rvert}
    \frac{
        \binom{
            j-1
        }{
            k-1
        }
    }{
        \binom{
            N \cdot \lvert \mathrm{procs} \rvert
        }{
            k
        }
    }
    \frac{
        T^*_p
    }{
        T_{p,j,n}
    }
\end{equation}
Here $\mathrm{procs}$ is the set of resource counts over which the experiments
can be performed. For example, if there are 128 hardware cores, $\mathrm{procs}
= {1, 2, 4, 8, 16, 32, 64, 128}$ processes or threads.

\subsubsection*{$\mathbf{efficiency_n@k}$}

To further understand the parallel performance of the generated code, we define
the $\mathrm{efficiency}_n@k$ metric. This metric measures the expected best
performance efficiency (speedup per process or thread) if the model is given
$k$ attempts to generate the code.  This is easily defined by modifying
\Cref{eq:speedup-k} to divide by $n$ as shown in \Cref{eq:efficiency-k}. The possible values of this
metric range between 0 and 1.0, with 1.0 representing a model that generates code
that scales perfectly with the number of processes or threads. This metric is useful for
understanding how well the generated code makes use of parallel resources. In
addition to $\mathrm{efficiency}_n@k$, we also define
$\mathrm{efficiency}_{\mathrm{max}}@k$ in the same fashion as
~\Cref{eq:speedup-max-k}.
\begin{equation}\label{eq:efficiency-k}
    \textrm{efficiency}_n@k =
    \frac{1}{\lvert P\rvert}
    \sum_{p\in P}
    \sum_{j=1}^N
    \frac{
        \binom{
            j-1
        }{
            k-1
        }
    }{
        \binom{
            N
        }{
            k
        }
    }
    \frac{
        T^*_p
    }{
        n \cdot T_{p,j,n}
    }
\end{equation}

Even though we explore parallel code generation in this paper, these metrics
can be used to consider the performance of sequential code generation as well.
For example, examining $\mathrm{speedup}_1@k$ for the HumanEval, MBPP, or
DS-1000 benchmarks will lead to a better understanding of how efficient the
generated Python code is compared to a human created baseline. Additionally,
both performance metrics could be modified to be parameterized by problem size
instead of number of processes/threads in order to study the computational
complexity of the generated code.

\section{Experimental Setup}
\label{sec:experimental-setup}
This section describes how we generate outputs using each of the LLMs
(Section~\ref{sec:models}) and the prompts in \generatebench{}, and how we
evaluated the generated code using the \generatebench{} test harness.

\subsection{LLM Inference: Generating Code Output}
\label{sec:setup-llm-inference}

To generate outputs with the open-source models, we use the HuggingFace
library~\cite{huggingface} with PyTorch~\cite{paszke2019pytorch} as the backend
to load the LLM weights and use them for inference. Specifically, we create a
PyTorch Dataset object that wraps the set of prompts and we pass this as input
to a Huggingface Pipeline object, which then runs the models in inference mode
and generates the outputs. We do these runs on a single NVIDIA A100 80GB GPU
using 16-bit floating point precision. Since the prompt workloads are fairly
regular, we get the best inference performance for larger batch sizes. So for
each model, we use the largest batch size that fits in GPU memory. To generate
the GPT-3.5 and GPT-4 outputs we use the OpenAI API~\cite{openai-api} via
OpenAI's Python client~\cite{openai-api-python}.

For all of the tasks, we use nucleus sampling with a value of $p=0.95$.
Additionally, we limit the maximum number of new tokens generated to 1024. We
experimentally found this to be long enough for all of the tasks to be
completed, but short enough to limit long, repetitive outputs. Using this
configuration, we create two sets of outputs for each model: one with 20
samples per prompt and a temperature of 0.2, and the other with 200 samples per
prompt and a temperature of 0.8. The former is used to calculate the metrics at
$k=1$ (such as pass@$1$) and the latter for larger values of $k$. This is in
line with the generation configurations in related
literature~\cite{li2023starcoder, roziere2023code}. Note that we exclude the
evaluation of GPT-3.5 and GPT-4 with 200 samples per prompt and a temperature
of 0.8 due to the high monetary cost of generating these outputs.

\subsection{Evaluating the Generated Code}\label{sec:setup-evaluation}

To evaluate the generated code, we use the \generatebench{} test harness. The
test harness is a set of scripts that compile and run the generated code using
manually written test drivers for each problem. The scripts handle recording
the compile status, correctness, and execution time of the generated code.

To compile the generated code, we use the GNU Compiler Collection (GCC) version
9.4.0. For serial, OpenMP, and Kokkos versions, we use GCC as the primary compiler,
whereas we use it as the backend to the respective frontend compiler for the
other models (i.e. the backend compiler to {\it mpicxx}). All compilations use
the flags \texttt{-O3 -std=c++17} and the OpenMP tasks add the
\texttt{-fopenmp} flag. We use version 4.1.0 of Kokkos, and the {\it threads}
execution space, which uses C++ threads for parallelism. MPI codes are compiled
with OpenMPI version 4.1.1. CUDA programs are compiled with {\it nvcc} and
CUDA version 12.1.1. Likewise, HIP programs are compiled with {\it hipcc} and
ROCm version 5.7.0. 

Before compiling an output, the prompt and generated code are written to a
header file that is included by the driver script for that task. Once compiled,
the generated binary is run by the test harness.  The test harness checks if
the generated code produces the same results as the sequential baseline.  The
sequential baselines are handwritten, optimal implementations of the prompt
that are used to test correctness and to calculate the performance metrics (see
\Cref{sec:metrics-performance}).  Additionally, a code can be labeled as
incorrect for the following reasons:
\begin{itemize}[leftmargin=*]
\item The code does not compile or it takes longer than three minutes to run.
We choose the problem sizes for each prompt such that any reasonable
implementations execute in much less than three minutes.
\item The code does not use its respective parallel programming model. For
example, if the model generates a sequential implementation rather than using
OpenMP when prompted to do so, it is labeled as incorrect. We utilize several
string matching criteria to implement this check.
\end{itemize}
The output of the program includes the result of the correctness check of the
generated code, the average runtime of the generated code, and that of the
sequential baseline over ten runs.  We use the default timer for each execution
model to measure its run time.
 
The CPU runs are conducted on an AMD EPYC 7763, 2.45 GHz CPU with 64 physical
cores and 512 GB of RAM. We run with $1, 2, 4, \ldots, 32$ threads for OpenMP
and Kokkos. For MPI, we run with $1, 2, 4, \ldots, 512$ processes across
multiple nodes with one process per physical core. For MPI+OpenMP we run on 1, 2,
3, and 4 nodes with 1 process per node and $1, 2, 4, \ldots, 64$ threads per
node. The CUDA runs are completed on an NVIDIA A100 80GB GPU and the AMD runs
on an AMD MI50 GPU. Kernels are launched with the number of threads indicated
in the prompt text (i.e. {\it at least as many threads as values in the
array}).

\section{Evaluation Results}
\label{sec:results}
We now present detailed results from evaluating the LLMs described in
\Cref{sec:models} using the \generatebench{} prompts and test harness.

\subsection{Experiment 1: Parallel Code Generation}
\label{sec:results-benchmark}

\begin{RQcallout}
    {\bf RQ1 }{\it \RQone{}}
\end{RQcallout}

\begin{figure}[h]
    \centering
    \includegraphics[width=\linewidth]{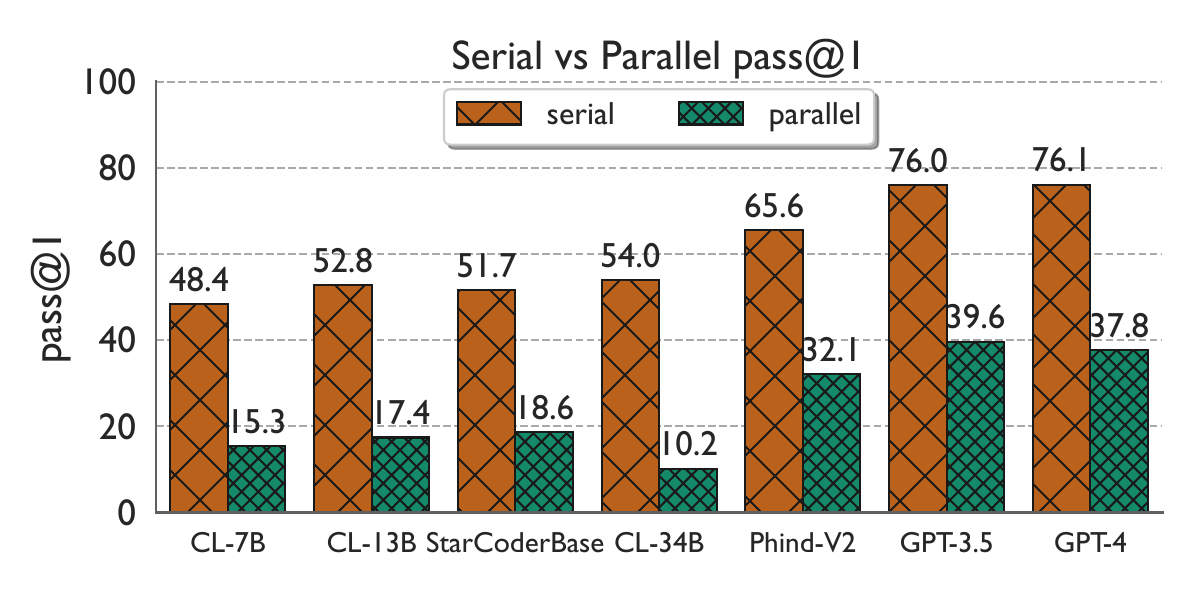}
    \caption{Each LLM's pass@1 score over \generatebench{}. All of the LLMs
score significantly worse in generating parallel code than serial code.  }
    \label{fig:pass1}
\end{figure}

To evaluate the correctness of the code generated by the LLMs we first look at
the pass@1 scores over \generatebench{}. \Cref{fig:pass1} shows the pass@1
score for each LLM for generating the serial code versus the average over
the six parallel execution models. As defined in \Cref{eq:pass-k}, these values
are aggregated over all the prompts including problem types and execution
models.  Notably, all of the LLMs score significantly worse for parallel code
generation than they do for serial code generation. The best performing models,
GPT-3.5 and GPT-4, both achieve \tweakedsim 76 pass@1 on the serial prompts. This
is a strong score in the context of other benchmarks, such as HumanEval, where
GPT-4 gets 84.1 (see \Cref{tab:models}). Despite the strong serial scores,
GPT-3.5 and GPT-4 only achieve 39.6 and 37.8 pass@1, respectively, on the
parallel prompts.

The open-source models show a significant decrease in performance for parallel
code generation with all of them except Phind-V2 (Phind-CodeLlama-V2) scoring
between 10.2 and 18.6.  Phind-V2 does much better than the other open-source
models, achieving 32 pass@1 on the parallel prompts. This suggests that further
fine-tuning of the open-source code models can improve their performance on
parallel code generation. Additionally, it is significant that an open-source
model performs near to the closed-source models on parallel code generation.
Open-source models are more accessible and, thus, having a strong open-source
model for parallel code generation would be beneficial to the community.

Another interesting trend we observe in \Cref{fig:pass1} is that CodeLlama-34B
and GPT-4 both score worse than their smaller counterparts on parallel code
generation. The reasons for this decrease in performance are not immediately
obvious. However, we observe that CodeLlama-34B and GPT-4 often generate the
same output for a given prompt for most or all of the 20 samples. This is due to
the larger models being more ``confident'' in their outputs, but this can have
an adverse effect on the pass@1 score when the output is incorrect.

Ultimately, the closed-source models are better than the open-source models at
parallel code generation. Interestingly, GPT-3.5 beats GPT-4 on the parallel
prompts by almost 2 percentage points, suggesting it may be better suited for
parallel code generation tasks. This is interesting since GPT-4 is bigger and
newer than GPT-3.5 and generally obtains better results on other code and
natural language benchmarks. Amongst the open-source models, Phind-V2 has the
best results, but still lags behind the closed-source models by almost 8
percentage points.

\begin{figure}[h]
    \centering
    \includegraphics[width=\linewidth]{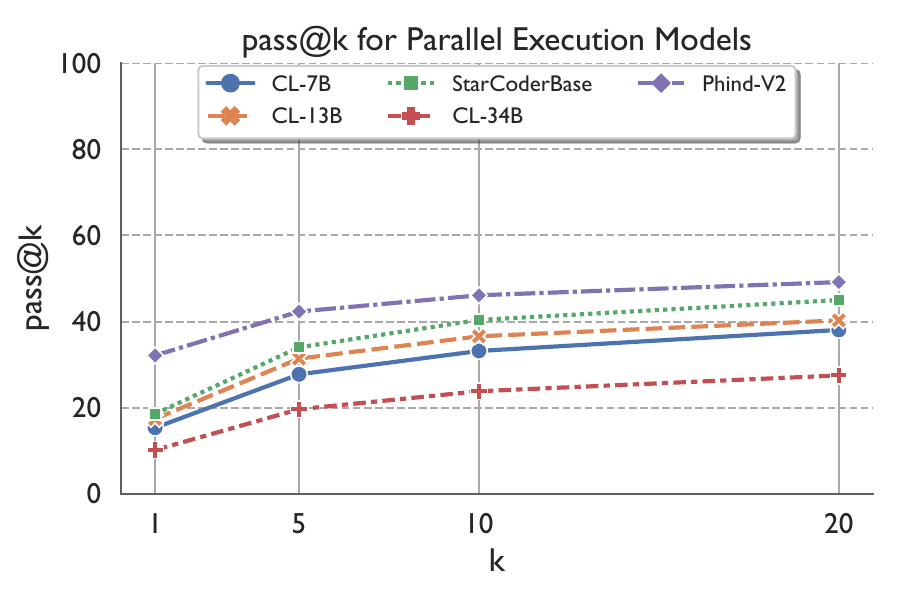}
    \caption{The pass@k for various values of k. The relative order of the LLMs
        is the same for all values of k with Phind-V2 leading the group.}
    \label{fig:pass-by-k}
\end{figure}

In addition to pass@1 it is also useful to consider pass@$k$ for $k>1$ to
understand how the LLMs perform provided more attempts at a problem.
Figure~\ref{fig:pass-by-k} shows the pass@$k$ for each LLM for $k=1,5,10,20$
with 200 samples and a temperature of 0.8 for $k\ne1$. The GPT models are
omitted for $k>1$ due to the monetary cost of generating a large number of
samples with these models. We observe the same relative ordering as in
\Cref{fig:pass1} is maintained for all values of $k$ with Phind-V2 leading the
open-source LLMs. At $k=20$ Phind-V2 achieves a pass@$k$ of 46 meaning that on
average it is able to generate a correct answer to one of the parallel prompts
in 20 attempts 46\% of the time. The scores of each LLM improving with an increase in $k$ is
expected due to the nature of the pass@$k$ metric. The fact that each LLM
begins to plateau suggests that there is an upper limit to their ability to
generate correct parallel code and giving them more attempts does not
significantly improve their performance.

\begin{RQcallout}
    {\bf RQ2 }{\it \RQtwo{}}
\end{RQcallout}

\subsubsection{\bf Breakdowns by Execution Models}

\begin{figure}[h]
    \centering
    \includegraphics[width=\linewidth]{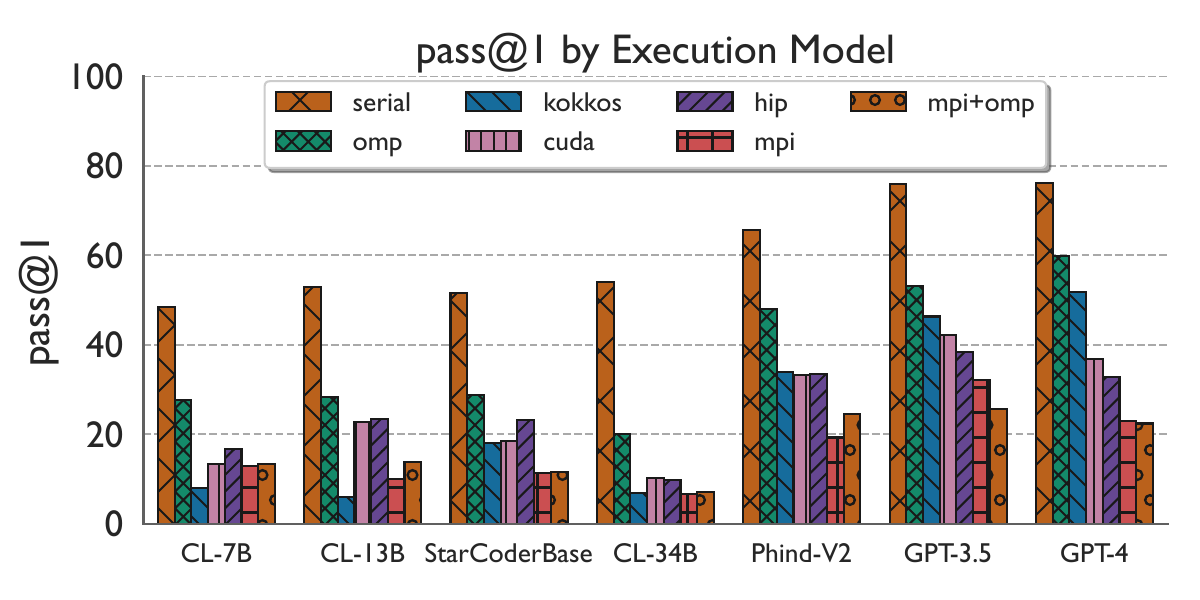}
    \caption{pass@1 for each execution model. The LLMs generally follow the same
        distribution of scores across the execution models: serial (best),
        OpenMP, CUDA/HIP, and MPI/MPI+OpenMP (worst) with Kokkos varying between
        LLMs.}
    \label{fig:pass1-by-execution-model}
 \end{figure}

We further break down the pass@1 results by each execution model in
\Cref{fig:pass1-by-execution-model}. From this data  we observe that every LLM
follows a similar distribution of scores across the execution models: serial
(best), OpenMP, CUDA/HIP, and MPI/MPI+OpenMP (worst) with Kokkos varying
between LLMs.

\begin{figure*}[t]
    \centering
    \includegraphics[width=0.95\linewidth]{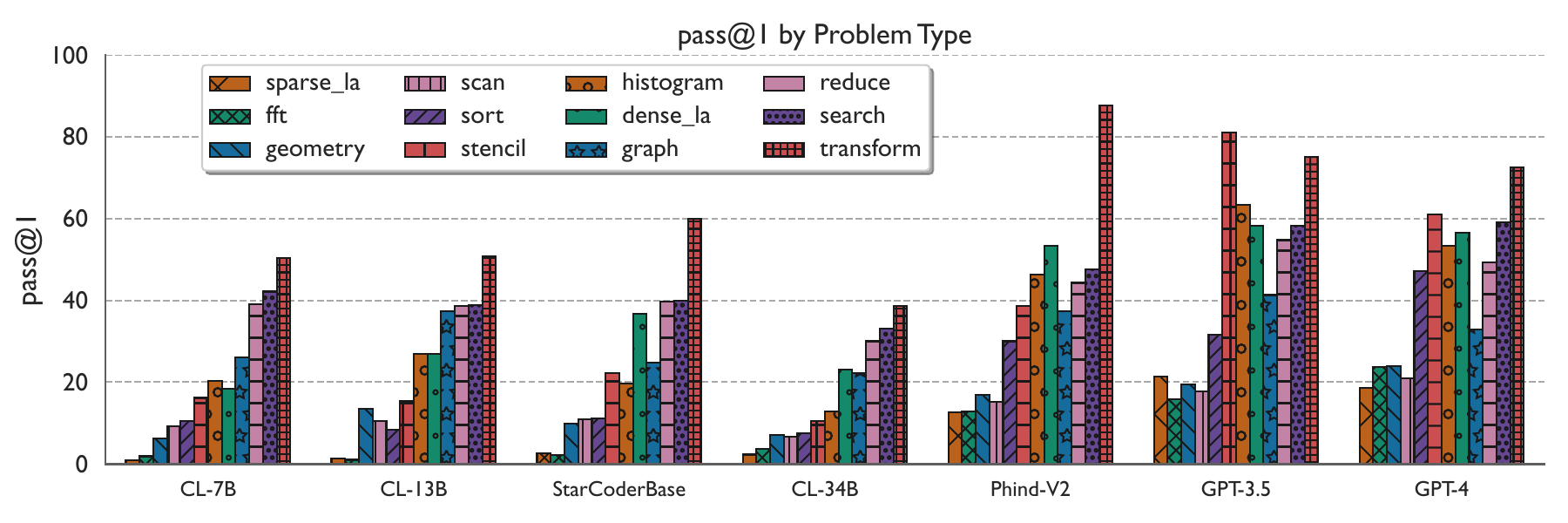}
    \caption{pass@1 for each problem type. The LLMs are best at
        transform problems, while they are worst at sparse linear
        algebra problems.}
    \label{fig:pass1-by-problem-type}
\end{figure*}

The pass@1 of LLMs being better with OpenMP than other parallel execution models
is likely due to the fact that OpenMP code is the most similar to serial code. For many
problems it only requires adding an OpenMP pragma, and occasionally a reduction
clause. GPT-4 gets nearly as many OpenMP problems correct as serial problems,
with an OpenMP pass@1 of 60 vs a 76 serial pass@1. The other top LLMs, GPT-3.5
and Phind-V2, are also nearly as efficient on OpenMP problems as serial
problems. StarCoderBase and the CodeLlama models have a larger gap between
their serial and OpenMP pass@1 scores, but still have better results on OpenMP
than the other parallel execution models.

With the larger LLMs, Kokkos is consistently just behind OpenMP in its pass@1
results. Like OpenMP, Kokkos is a shared memory parallel programming model that
relies mostly on high-level abstract constructs to parallelize code. These
high-level abstractions make it simpler for the LLM to translate the prompt
text to code. The smaller LLMs struggle with Kokkos, likely due to the fact
that Kokkos is more verbose than OpenMP and is more niche than the other
parallel execution models leading to less inclusion in their training data.
With fewer Kokkos examples in the dataset the smaller LLMs likely struggle to
learn how to model Kokkos code well.

Following Kokkos, we observe that all the LLMs are next most efficient for
CUDA/HIP. These two always have a similar pass@1 score, which is likely due to the
similarity of CUDA and HIP. All of the open-source LLMs have a slightly better
pass@1 with HIP than CUDA, while the closed-source LLMs are slightly better
with CUDA than HIP. CUDA/HIP kernels are more complex than OpenMP and Kokkos,
but the parallelism is intrinsic to the kernel making it easier than MPI, since
the LLM does not need to reason about large changes to the underlying
algorithm.

MPI and MPI+OpenMP are generally the worst parallel execution models for all
the LLMs (except for CodeLlama 7B and 13B where they are second and third
worst).  Compared to the other execution models in our testing, MPI
implementations often differ the most from their sequential counterparts. This
complexity makes it difficult for the LLMs to generate correct MPI code. Based
on the results for all the execution models, we hypothesize that this trend
generalizes to all parallel execution models: the more different a parallel
programming model's code is from the corresponding serial code, the more difficult it is for the
LLMs to generate correct code in that programming model.

\subsubsection{\bf Breakdowns by Problem Types}

In addition to execution models it is also important to understand what types
of computational problems LLMs struggle to parallelize.
\Cref{fig:pass1-by-problem-type} shows the pass@1 score for each problem type
across all the LLMs. As a general trend, we observe that all LLMs are better at
generating parallel solutions for structured, dense problems and worse for
unstructured, sparse problems.

All of the LLMs get their best pass@1 scores for transform problems with the
exception of GPT-3.5 where it is the second best. Transform problems are the
simplest as they are completely data parallel. In addition to transform, all of
the LLMs generally score well on reduction and search. These are also fairly
simple to parallelize as searching requires little to no communication and
reductions are often offered as high-level constructs in parallel programming
models.

Phind-V2 and the GPT LLMs score well on stencil, histogram, and dense linear
algebra problems. These problems are all structured and dense, which makes them
easier for the LLMs to parallelize. These three problems are in the middle of
the group for StarCoderBase and the CodeLlama LLMs coming after transform,
search, and reduce. This suggests that the larger LLMs are better at
parallelizing these types of problems. Interestingly, StarCoderBase and the
CodeLlama LLMs all have graph problems in their top four to five problem types,
which is not the case for Phind-V2 and the GPTs.

The bottom five problem types for all of the LLMs are sparse linear algebra,
scan, fft, geometry, and sort. GPT-4 is the exception with graph instead of
sort as the fifth-worst problem type. Sparse linear algebra is generally the
worst problem type, which is likely due to the difficulty in parallelizing
sparse computations. FFT and geometry problems are also generally more
difficult to parallelize so it readily follows that the LLMs would struggle
with them. The sorting and scan results are more surprising. Parallel
implementations for sort and scan are well known and certain execution models
like OpenMP and MPI even offer high-level abstractions for scan. 

\begin{figure}[h]
    \centering
    \includegraphics[width=\linewidth]{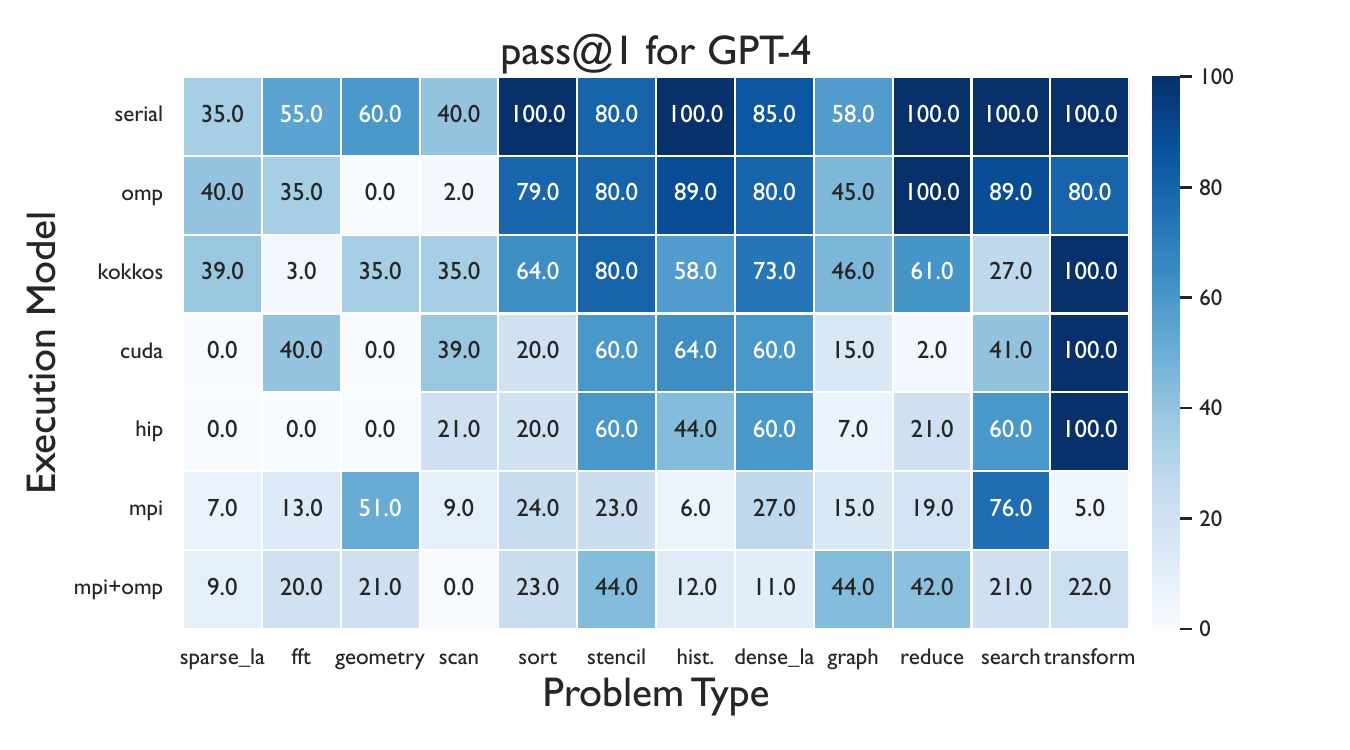}
    \caption{pass@1 for GPT-4 across all execution models and problem types.
        GPT-4 excels with the Kokkos and OpenMP execution models, while getting
        more problems correct for transform, search, and reduce problems.}
    \label{fig:pass1-gpt4-heatmap}
\end{figure}

\Cref{fig:pass1-gpt4-heatmap} provides an even more detailed view of the pass@1
metric across both execution models and problem types for GPT-4. We see the
same trends as in \Cref{fig:pass1-by-execution-model,fig:pass1-by-problem-type}
for GPT-4, however, we can also see where certain trends do not hold. For
example, despite being the best LLM for search problems and the best LLM at
Kokkos, GPT-4 does not do well on Kokkos search problems. We also see that MPI
and MPI+OpenMP scores on a particular problem type are not always the same.
This suggests that the model has difficulty dealing with these dual execution
models.

\vspace{2em}
\begin{RQcallout}
    {\bf RQ3 }{\it \RQthree{}}
\end{RQcallout}

\subsubsection{\bf Speedup and Efficiency}
When writing parallel code, it is important to consider performance in
addition to correctness. A parallel implementation that is correct, but makes
inefficient use of resources is not useful in practice. Hence, we compare
the speedup$_n$@k and efficiency$_n$@k metrics for each LLM.

\begin{figure}[h]
    \centering
    \includegraphics[width=\linewidth]{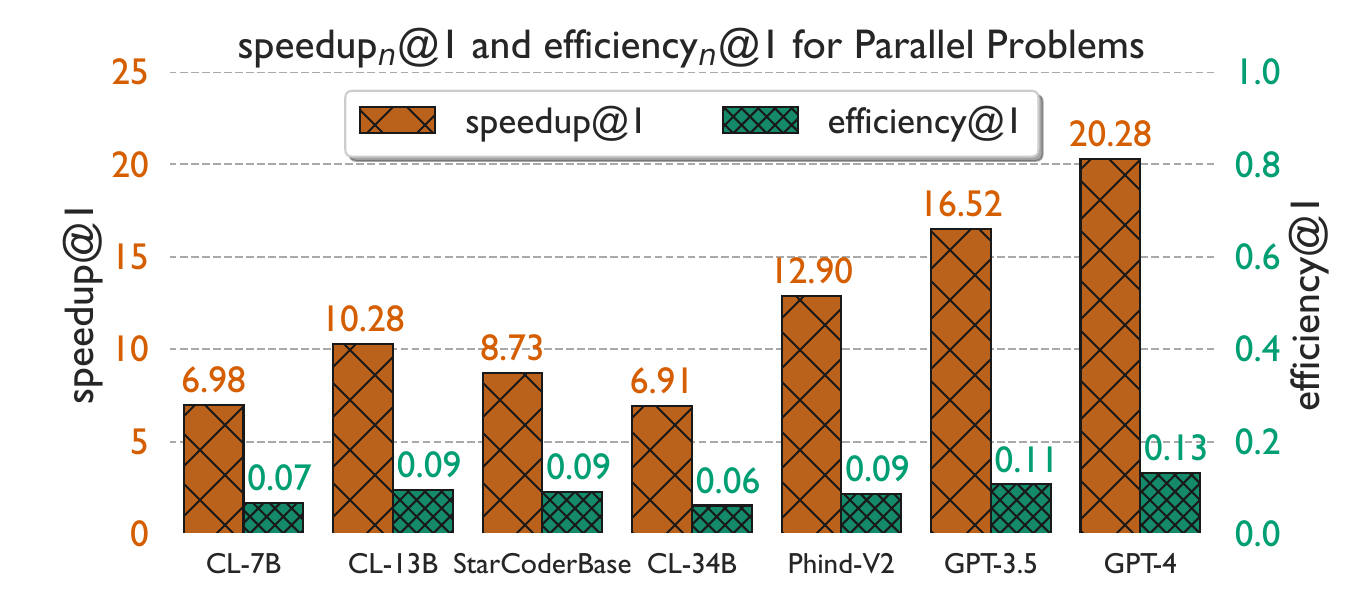}
    \caption{$\mathrm{speedup}_n@1$ and $\mathrm{efficiency}_n@1$ for parallel
        prompts. Results are shown for $n=32\,\mathrm{threads}$ for OpenMP and
        Kokkos, $n=512\,\mathrm{ranks}$ for MPI, and $n=(4\,\mathrm{ranks})
        \times (64\,\mathrm{threads})$ for MPI+OpenMP. For CUDA/HIP $n$ is set
        to the number of kernel threads, which varies across prompts.
        \protect\footnotemark[1]}
    \label{fig:speedup1-and-efficiency1-by-model}
\end{figure}

\begin{figure*}[t]
    \centering
    \includegraphics[width=0.3\linewidth]{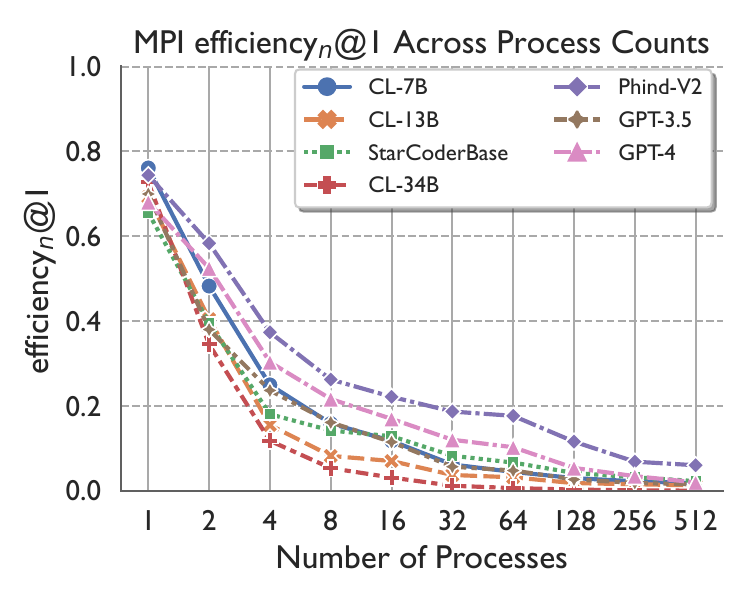}
    \includegraphics[width=0.3\linewidth]{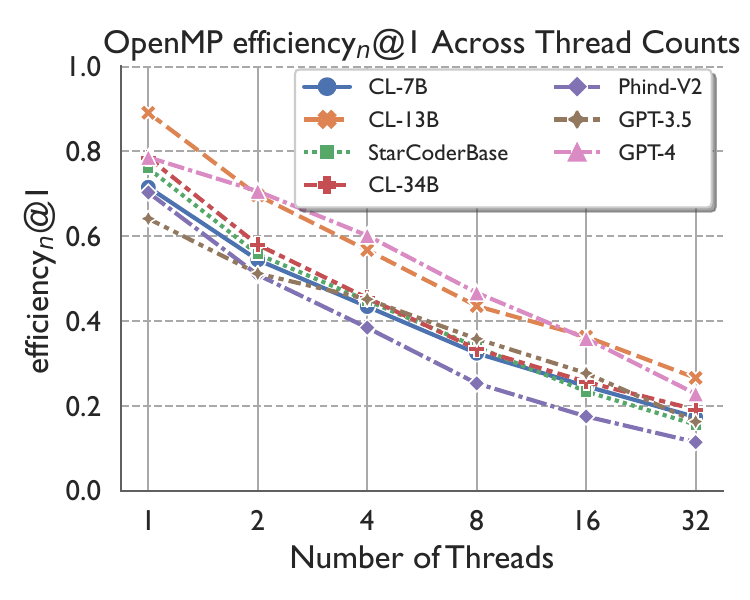}
    \includegraphics[width=0.3\linewidth]{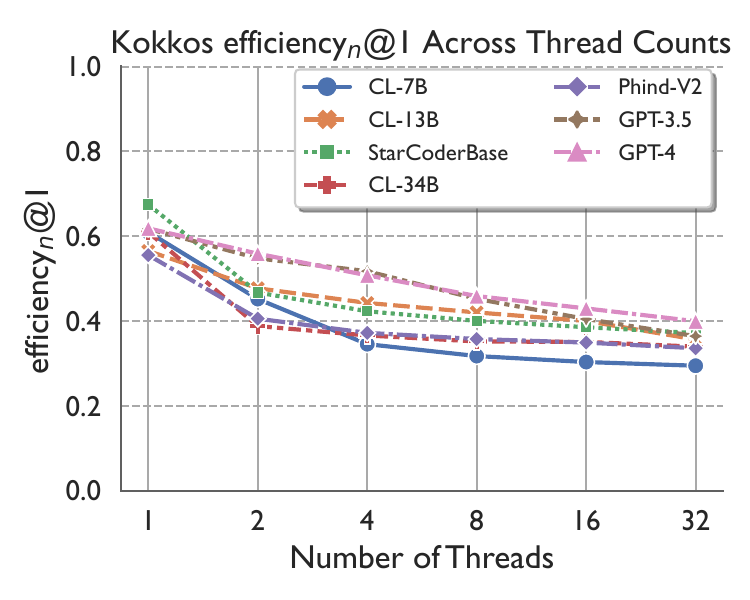}
    \caption{efficiency@1 for MPI (left), OpenMP (middle), and Kokkos (right) 
        prompts across rank and thread counts. Phind-V2 is most efficient for
        MPI prompts, but is one of the least efficient for OpenMP and Kokkos.
        GPT-4 is the most efficient for OpenMP and Kokkos prompts.
        \protect\footnotemark[1]}
    \label{fig:efficiency-by-model-mpi-omp-kokkos}
\end{figure*}

\Cref{fig:speedup1-and-efficiency1-by-model} shows the speedup$_n$@1 and
efficiency$_n$@1 scores for each LLM, averaged across the parallel execution models. For comparison,
we use the highest value of $n$ for each execution model that we ran in our
experimentation: $n=32\,\mathrm{threads}$ for OpenMP and Kokkos,
$n=512\,\mathrm{processes}$ for MPI, and $n=(4\,\mathrm{processes}) \times
(64\,\mathrm{threads})$ for MPI+OpenMP. For CUDA/HIP, $n$ is set to the number
of kernel threads, which varies across prompts.\footnote[1]{{\it Search}
problems are omitted from speedup$_n$@k and efficiency$_n$@k results due to
their high super-linear speedups preventing a meaningful analysis of the
performance results for other problem types.}

In \Cref{fig:speedup1-and-efficiency1-by-model}, we see a trend similar to the pass@1 scores in \Cref{fig:pass1}, with the GPT
models scoring the highest and the CodeLlama models scoring the lowest. Despite
GPT-3.5 having the highest pass@1 for parallel prompts, GPT-4 has the highest
speedup$_n$@1 for all parallel execution models at 20.28. This means that on
average GPT-4's parallel code achieves a 20.28x speedup over the sequential
baseline. To help interpret this result, we also show the efficiency$_n$@1 for each
LLM for the parallel prompts in \Cref{fig:speedup1-and-efficiency1-by-model}.
From this we see that none of the LLMs use parallel resources efficiently. The
best efficiency$_n$@1 is 0.13 for GPT-4 meaning that on average GPT-4's
parallel code achieves 13\% of the maximum possible speedup. CodeLlama-34B has
the worst efficiency$_n$@1 at 0.06. From the results in
\Cref{fig:speedup1-and-efficiency1-by-model} we can conclude that the parallel
code produced by LLMs is generally inefficient even when correct.

It is also important to consider how efficiency$_n$@1 varies across $n$.
\Cref{fig:efficiency-by-model-mpi-omp-kokkos} compares the efficiency$_n$@1
curves for MPI, OpenMP, and Kokkos. We see Phind-V2 is the most efficient at
MPI prompts, while the least efficient at OpenMP and second to least for
Kokkos.  GPT-4 produces the most efficient code on average as it is one of the
top two most efficient for all three execution models. All of the models start
with better efficiency$_n$@1 for OpenMP than Kokkos, but rapidly decline
towards an efficiency$_n$@1 of \tweakedsim $0.2$. On the other hand, the Kokkos
efficiency$_n$@1 values stay roughly consistent across $n$, showing efficient
use of threads.

\begin{figure}[h]
    \centering
    \includegraphics[width=\linewidth]{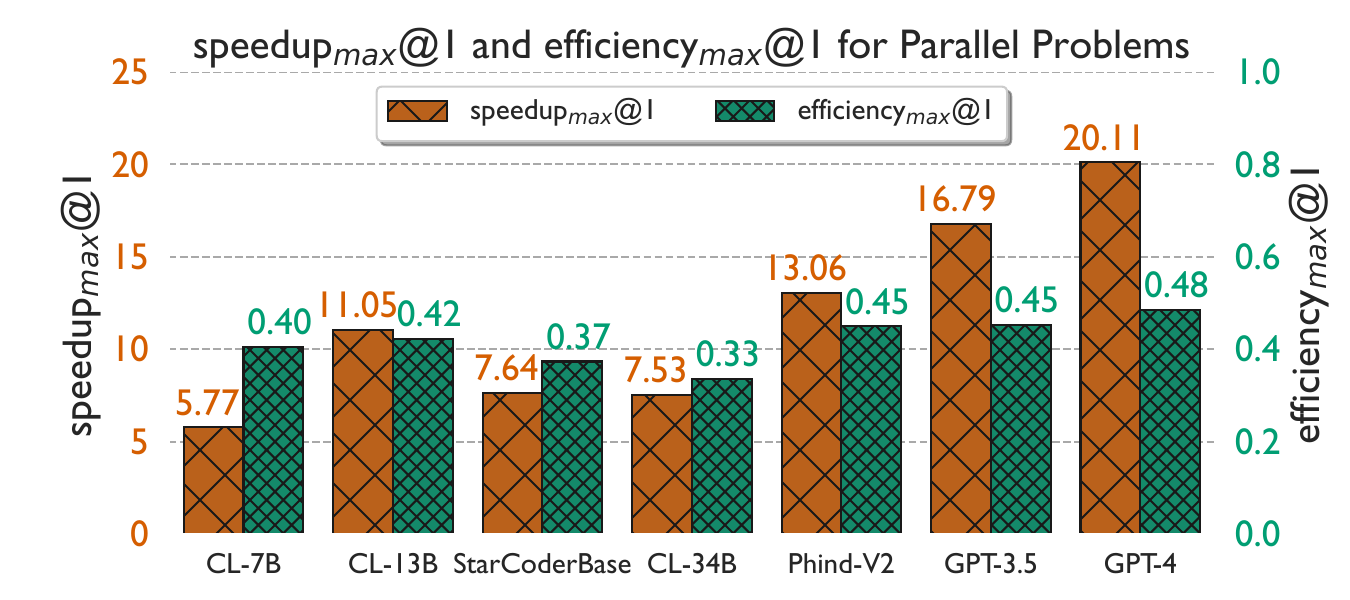}
    \caption{The expected max speedup and efficiency across all resource counts
    $n$. \label{fig:speedup1_max-and-efficiency1_max-by-model}}
\end{figure}

\Cref{fig:speedup1_max-and-efficiency1_max-by-model} further shows the expected
maximum speedup and efficiency across all resource counts $n$. We see the same
trends as in \Cref{fig:speedup1-and-efficiency1-by-model} with the speedups at
similar values and the efficiencies higher. This is likely due to a number of
the generated code samples plateauing at a certain $n$, so choosing a smaller
$n$ can give a better efficiency with the same speedup.

\subsection{Experiment 2: Parallel Code Translation}
\label{sec:results-translation}

\begin{RQcallout}
    {\bf RQ4 }{\it \RQfour{}}
\end{RQcallout}

\begin{figure}[h]
    \centering
    \includegraphics[width=\linewidth]{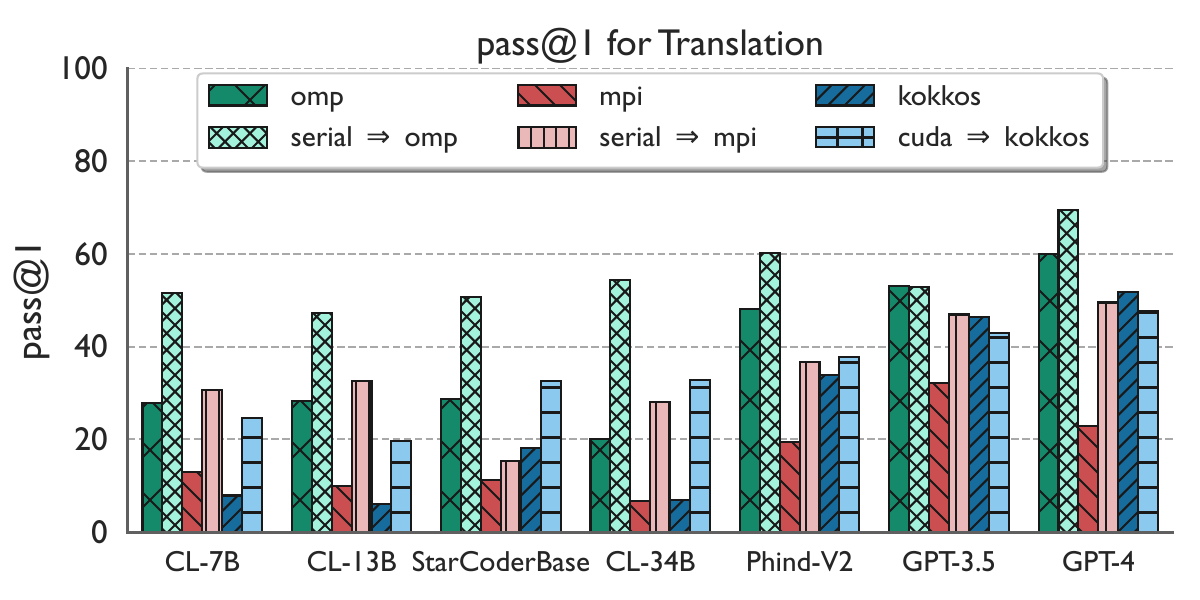}
    \caption{pass@1 for each LLM when translating serial to OpenMP, serial to
        MPI, and CUDA to Kokkos compared to the pass@1 score for generating code
        in the destination execution model. The smaller LLMs see a significant
        improvement when shown an example correct implementation.}
    \label{fig:pass1-translate}
\end{figure}

In addition to generating parallel code from scratch, we also evaluate the LLMs
ability to translate between execution models (see
\Cref{sec:setup-code-translation}). \Cref{fig:pass1-translate} shows the pass@1
scores for each LLM for translating serial to OpenMP, serial to MPI, and CUDA
to Kokkos. We also include the generation pass@1 scores from
\Cref{fig:pass1-by-execution-model} for each LLM for OpenMP, MPI, and Kokkos.

Several LLMs score significantly better when given a correct example
implementation in a different execution model i.e.~translation. All LLMs, except
for GPT-3.5, have a higher pass@1 score for translating to OpenMP than they do
for generating OpenMP code from scratch. We observe that the LLMs are able to
correctly parallelize the provided serial code with OpenMP. A similar trend
emerges with the serial to MPI translation. All of the LLMs score better when
translating serial code to MPI than they do when generating MPI code from
scratch. Likewise, all of the LLMs see an improvement translating from CUDA to
Kokkos over native Kokkos generation with the exception of the GPT models. 

It is expected that the pass@1 scores would either increase or stay the same,
since the LLM is given more information during translation than when generating
code from scratch. It is surprising, however, the magnitude of improvement that
the smaller LLMs experience. For example, CodeLlama-7B has a pass@1 of 20 for
generating OpenMP code from scratch, but a pass@1 of 52 for translating serial
code to OpenMP. This suggests that providing LLMs with correct implementations
can improve their ability to generate correct parallel code.

\subsubsection{\bf Speedup and Efficiency}
\begin{figure}[h]
    \centering
    \includegraphics[width=\linewidth]{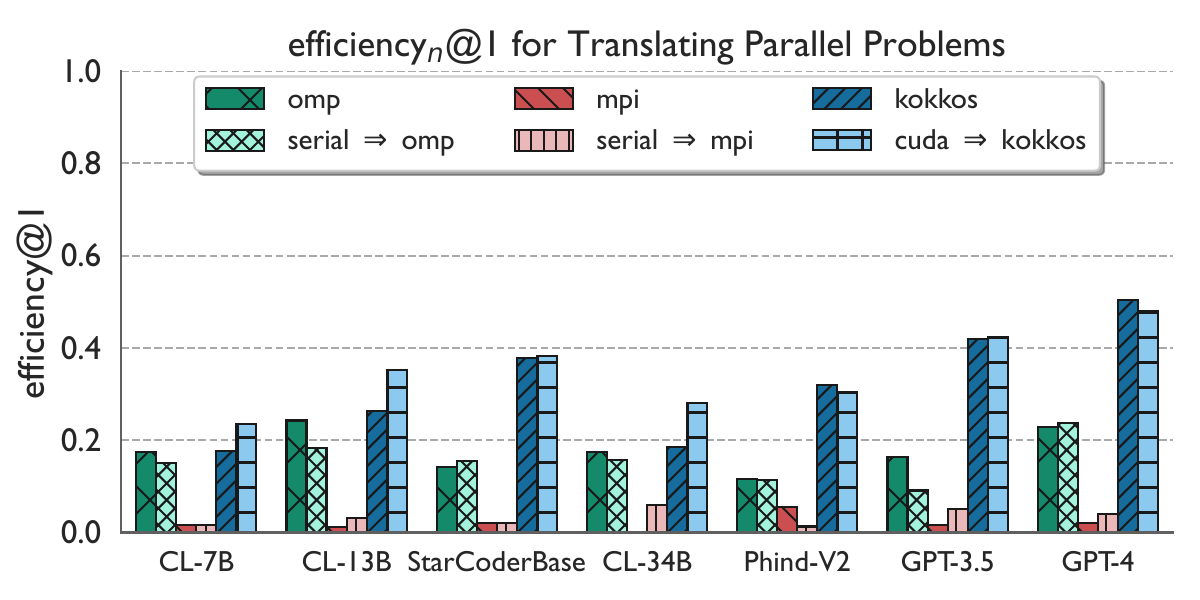}
    \caption{efficiency@1 translation scores compared to generation scores.
        The LLMs generally score similarly for translation and generation.\protect\footnotemark[1]}
    \label{fig:efficiency1-translate}
\end{figure}

While translating between execution models improves the pass@1 score it does
not generally improve the performance of the generated code as shown in
\Cref{fig:efficiency1-translate}. Most LLMs see a similar efficiency$_n$@1 for
OpenMP, MPI, and Kokkos whether generating from scratch or translating between
execution models. A number of LLMs actually perform worse when translating from
serial to OpenMP. 

\begin{figure}[h]
    \centering
    \includegraphics[width=\linewidth]{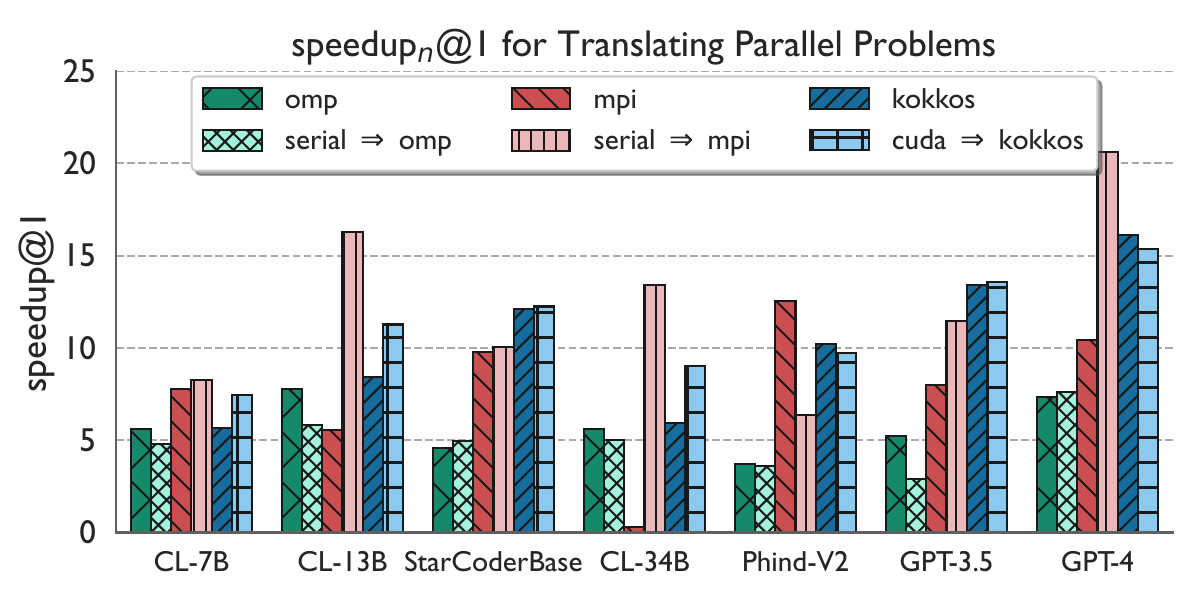}
    \caption{speedup@1 translation scores compared to generation scores.
        The LLMs generally perform similarly for translation and generation with the exception of MPI.\protect\footnotemark[1]}
    \label{fig:speedup1-translate}
\end{figure}

We observe similar trends with OpenMP and Kokkos for speedup$_n$@1 as shown in
\Cref{fig:speedup1-translate}. The LLMs generally perform similarly for
translation and generation. The exception is MPI where CodeLlama-13B,
CodeLlama-34B, and GPT-4 all get significantly better speedup$_n$@1 when
translating from serial to MPI code. From the results in
\Cref{fig:pass1-translate,fig:speedup1-translate,fig:efficiency1-translate} we
conclude that providing LLMs with correct implementations in one execution
model helps them generate correct code in another execution model, but does not
necessarily improve the performance of the generated code.

\section{Conclusion}
\label{sec:conclusion}
In this paper, we proposed a \generatebenchfull{} (\generatebench{}) benchmark for
evaluating the ability of LLMs to generate parallel code. We additionally
introduced two novel metrics for evaluating the runtime performance and scaling
behavior of the generated parallel code. Using \generatebench{} and these
metrics, we have evaluated the ability of state-of-the-art open- and
closed-source LLMs to generate parallel code. We find that LLMs are
significantly worse at generating parallel code than they are at generating
serial code. In particular, we find that LLMs struggle most with MPI code and
sparse, unstructured problems. Further, we observe that closed-source models
outperform all the open-source models we tested, and that even when LLMs
generate correct parallel code, it is often not performant or scalable.
Providing correct implementations in one execution model (i.e. serial) helps
LLMs generate correct parallel code, but does not necessarily improve the
performance or scalability of the generated parallel code.

The poor performance of LLMs on \generatebench{} indicates that further efforts
are necessary to improve the ability of LLMs to model parallel code and/or
create new LLMs that are specialized for parallel code generation. These LLMs
will need to improve both the correctness and runtime performance of their
outputs. Benchmarks, such as \generatebench{}, are vital to creating and
improving LLMs for parallel code generation. By iterating on \generatebench{}
and the metrics we have proposed, we can continue to improve the ability of
LLMs in this domain and create state-of-the-art open-source LLMs for different parallel
code development tasks.

\begin{acks}
This material is based upon work supported in part by the National Science
Foundation under Grant No.~2047120, and by the National Science Foundation
Graduate Research Fellowship Program under Grant No.~DGE~2236417. This research
used resources of the National Energy Research Scientific Computing Center, a
U.S.~Department of Energy Office of Science User Facility using NERSC award
DDR-ERCAP0025593. We spent \tweakedsim 80 dollars for the use of the paid API of
GPT-3.5 and GPT-4 for the evaluation in this paper.

\end{acks}

\bibliographystyle{ACM-Reference-Format}
\bibliography{bib/pssg,bib/cite}

\end{document}